\documentclass[letterpaper,10pt,conference]{ieeeconf}
\usepackage{color}
\usepackage[dvipsnames]{xcolor}
\usepackage{amsmath}
\usepackage{amssymb}
\usepackage{graphicx}
\usepackage{bm}
\usepackage{url}
\usepackage[font=footnotesize, skip=4pt]{caption}
\usepackage{algorithm,algorithmic}
\usepackage{subfig}
\usepackage{tikz}
\usetikzlibrary{bayesnet}

\newtheorem{assumption}{Assumption}

\IEEEoverridecommandlockouts

\makeatletter
\def\BState{\State\hskip-\ALG@thistlm}
\makeatother

\begin{document}

\title{\LARGE \bf A Bayesian Perspective on Residential\\ Demand Response Using Smart Meter Data}
\author{Datong Zhou, Maximilian Balandat, and Claire Tomlin\thanks{Datong Zhou is with the Department of Mechanical Engineering, University of California, Berkeley, USA. {\tt\footnotesize datong.zhou@berkeley.edu}}
\thanks{Maximilian Balandat and Claire Tomlin are with the Department of Electrical Engineering and Computer Sciences, University of California, Berkeley, USA.
{\tt\footnotesize [balandat,tomlin]@eecs.berkeley.edu}}%
\thanks{This work has been supported in part by the National Science Foundation under CPS:FORCES (CNS-1239166).}
}

\maketitle
\thispagestyle{empty}
\pagestyle{empty}

\begin{abstract}
The widespread deployment of Advanced Metering Infrastructure has made granular data of residential electricity consumption available on a large scale. Smart meters enable a two-way communication between residential customers and utilities. One field of research that relies on such granular consumption data is Residential Demand Response, where individual users are incentivized to temporarily reduce their consumption during periods of high marginal cost of electricity. To quantify the economic potential of Residential Demand Response, it is important to estimate the reductions during Demand Response hours, taking into account the heterogeneity of electricity users. In this paper, we incorporate latent variables representing behavioral archetypes of electricity users into the process of short-term load forecasting with Machine Learning methods, thereby differentiating between varying levels of energy consumption. The latent variables are constructed by fitting Conditional Mixture Models of Linear Regressions and Hidden Markov Models on smart meter readings of a Residential Demand Response program in the western United States. We observe a notable increase in the accuracy of short-term load forecasts compared to the case without latent variables. We then estimate the reductions during Demand Response events conditional on the latent variables, and discover a higher DR reduction among users with automated smart home devices compared to those without.
\end{abstract}


%
%


\section{Introduction}
\label{sec:Introduction}
Residential Demand Response (DR) is a novel data-driven service enabled by the large-scale deployment of Advanced Metering Infrastructure (AMI). By communicating a proxy of the marginal price of electricity to consumers, it is acknowledged that economic efficiency can be increased \cite{Commission:2010aa}. During times when the grid is strained, a DR provider, which serves as a mediating unit between residential electricity consumers and the DR market, bids reductions with respect to an expected consumption (baseline) into the wholesale electricity market. Different market regulators, including CAISO, have launched such pilot programs \cite{PJM:2014aa:CapPerf, CPUC:2015:E4728}. If the bid is cleared, the DR provider then prompts residential customers to temporarily reduce their consumption in exchange for a monetary reward proportional to the estimated reduction during DR times. As it is impossible to observe both the consumption conditional on DR-treatment and Non-DR-treatment, it becomes essential to estimate the counterfactual consumption, i.e. the consumption during DR times that would have been observed if no treatment had occurred. This is an application of the ``Fundamental Problem of Causal Inference" \cite{Holland:1986aa}, which states that it is impossible to observe more than one treatment on the same subject at one time.

For economic purposes, it is of cardinal importance for DR providers to bid the right amount of reductions into the wholesale electricity market, since penalties incur for negative shortfalls from the bidded capacity, and a suboptimal revenue would be recorded for a too modest bid. Assuming the bid is cleared, the major uncertainty is found to be the user behavior during DR times, i.e. the amount of reduction in response to the DR treatment. In \cite{Zhou:2016aa}, the authors find a positive correlation between the variability in consumption behavior and the magnitude of DR reduction, which suggests targeting variable households for a higher reduction yield.

In this paper, we analyze the heterogeneity in users' reduction behavior during DR times by using latent variables in statistical forecasting methods. This Bayesian perspective allows us to postulate the existence of behavioral archetypes of users, which govern the resulting and observable energy consumption. The latent variables are constructed in two ways: Firstly, we use a Conditional Gaussian Mixture Model (CGMM) of Linear Regressions, where the latent variable of a given data point is a vector of probabilities, with each component indicating the probability that the data point was generated by the corresponding mixture component. Secondly, we implement a Hidden Markov Model (HMM) whose hidden layer encodes hourly binary latent variables representing high and low levels of consumption, which in turn can be interpreted as an indicator for occupancy. The recommendation to DR providers is to prompt users only during hours of believed presence at home, thereby improving efficiency of targeting. Using this differentiation between different magnitudes of consumption, we observe a stark contrast in the estimated reduction between periods of high and low consumption.

In the extant literature, short-term load forecasting (STLF) has been extensively studied with different approaches and on different levels of aggregations of users, ranging from the individual level to city-wide predictions \cite{Sevlian:2014aa, Mirowski:2014aa}. Statistical time series models \cite{Arora:2014aa, Taylor:2007aa}, standard parametric regression models such as Ordinary Least Squares, Lasso- and Ridge-Regression \cite{Mirowski:2014aa}, and non-parametric methods including $k$-Nearest Neighbors, Support Vector Regression \cite{Elattar:2010aa}, and Neural Networks \cite{Edwards:2012aa} have been evaluated with respect to different metrics for accuracy. Widely explored Bayesian Methods for STLF are Gaussian Processes \cite{Lauret:2012aa}, Bayesian Neural Network approaches, e.g. for input selection problems \cite{Hippert:2010aa}, and Kalman-Filtering methods \cite{Al-Hamadi:2004aa} with Hybrid Neural Network extensions \cite{Guan:2013aa}. HMMs for STLF have been applied primarily for the purpose of occupancy detection \cite{Andersen:2014aa, Kleiminger:2014aa} and Nonintrusive Load Monitoring \cite{Parson:2011aa}. \cite{Han:2012aa} and \cite{Ai:2014aa} utilize occupancy information to increase the energy efficiency of building operation. To the best of our knowledge, CGMMs have not been investigated for STLF.

The contribution of this paper is two-fold: First, it aims to explore the potential for improvement in the prediction accuracy obtained by incorporating latent variable information from CGMMs and HMMs as an additional covariate into regression models. Second, it provides insights into the reduction behavior of users conditional on their latent states. Both aspects can help the DR provider make more informed bids into the wholesale market by targeting only the most susceptible users. The remainder of this paper is structured as follows: In Section \ref{sec:Forecasting_Methods}, we briefly outline classical Machine Learning (ML) methods used for STLF. Sections \ref{sec:Mixture_Models} and \ref{sec:HMMs} describe technical details of CGMMs and HMMs tailored to the specific needs of STLF, followed by Section \ref{sec:STLF}, which outlines the procedure of incorporating the estimated latent variables into STLF. Section \ref{sec:Reductions} presents a framework for estimating counterfactual consumption, which allows for the computation of the magnitude of the reduction of electricity consumption during DR hours. A case study on both semi-synthetic and observational data is presented in Section \ref{sec:Real_Data}. Chapter \ref{sec:Conclusion} concludes the paper.



%
%


\section{Forecasting Methods}
\label{sec:Forecasting_Methods}
In this section, we briefly describe well-established forecasting methods that we use in the remainder of this paper. Note, however, that a detailed description of these methods is outside the scope of this paper, and so we merely present these for completeness of the paper. The interested reader is referred to \cite{Zhou:2016aa} and the references therein.

\textit{Notation}: Let $Y \in \mathbb{R}^N$ denote a column vector of $N$ scalar outcomes $\lbrace y_1, \ldots, y_N \rbrace$, e.g. in our case electricity consumption, and $X \in \mathbb{R}^{N\times d}$ the design matrix whose $k$-th row represents the covariates $x_k \in \mathbb{R}^d$ associated with outcome $y_k$. Let $y$ and $x$ denote a generic outcome and its associated covariate vector, respectively.

\subsection{Ordinary Least Squares Regression}
\label{sec:LS_Regression}
Assuming a linear relationship between covariate-outcome pairs $(X, Y)$,
\begin{equation}\label{eq:X_y_linear}
Y = X w,
\end{equation}
the regression coefficients $w \in \mathbb{R}^d$ are estimated using Ordinary Least Squares Regression (OLS).

\subsection{K-Nearest Neighbors-Regression (KNN)}
\label{sec:KNNR}
Given a point in feature space $x$, the goal is to find the $k$ training points $x_1, \ldots, x_k$ that are closest in distance to $x$. We choose the commonly used Euclidian norm (though other choices can be justified) as a measure for distance in feature space. The prediction of the outcome $\hat{y}$ is the average of the outcomes of the $k$ nearest neighbors 
\begin{equation}
\hat{y} = \frac{1}{k}( y_1 + \ldots + y_k ).
\end{equation}
The number of neighbors $k$ for an optimal fit is found using common cross-validation techniques.

\subsection{Support Vector Regression}
\label{sec:SVR}
Support Vector Regression (SVR) solves the following optimization problem:
\begin{equation}\label{eq:SVR_opt}
\begin{aligned}
&\min_{w, b, \xi, \xi^{\ast}} \frac{1}{2} \|w\|^2 + C\sum_{i=1}^{N}(\xi_i + \xi_i^{\ast})\\
\text{s.t.}~ &y_i - w^{\top} \phi(x_i) - b \leq \epsilon +\xi_i,\\
&w^{\top} \phi(x_i) + b - y_i \leq \epsilon +\xi_i^{\ast},\\
&\xi_i, \xi_i^\ast \geq 0, \quad i \in \left[1, \ldots, N\right].
\end{aligned}
\end{equation}
In \eqref{eq:SVR_opt}, $\epsilon$ defines an error tube within which no penalty is associated, $\xi$ and $\xi^{\ast}$ denote slack variables that guarantee the existence of a solution for all $\epsilon$, $b$ is a real constant, $C$ is the regularization constant, $w$ are the regression coefficients to be estimated, and $\phi(\cdot)$ a map between the input space and a higher dimensional feature space. \eqref{eq:SVR_opt} is typically solved by transforming it into dual form, thereby avoiding the explicit calculation of $\phi(\cdot)$ with the so-called Kernel trick. We choose the commonly used Gaussian Kernel function.

\subsection{Decision Tree Regression (DT)}
\label{sec:DTR}
This non-parametric learning method finds decision rules that partition the feature space into up to $2^n$ pieces, where $n$ denotes the maximal depth of the tree. For a given iteration step, enumeration of all nodes and possible splitting scenarios (exhaustive search) yields a tuple $\theta^{\ast} = (j, t_m)$ that minimizes the sum of the ensuing child node impurities $G(\theta^{\ast}, m)$, where $j$ denotes the $j$-th feature and $m$ the $m$-th node of the tree. This is written as
\begin{subequations}
\begin{align}\label{eq:DT_optimization}
\theta^{\ast} &= \arg\min_{\theta} G(\theta, m), \\
\label{eq:DT_optimization_2} G(\theta, m) &= \frac{n_{\text{left}}^m}{N_m}H(Q_{\text{left}}(\theta)) + \frac{n_{\text{right}}^m}{N_m}H(Q_{\text{right}}(\theta)).
\end{align}
\end{subequations}
where $Q_{\text{left}}$ and $Q_{\text{right}}$ denote the set of covariate-outcome pairs belonging to the left and right child node of parent node $m$, respectively; and $n_{\text{left}}^m$ and $n_{\text{right}}^m$ denote their respective count. The impurity measure $H(\cdot)$ at a node minimizes the mean squared error
\begin{subequations}
\begin{align}\label{eq:impurity_MSE}
c(\cdot) &= \frac{1}{N(\cdot)} \sum_{i \in N(\cdot)} y_i, \\
\label{eq:impurity_MSE_2} H(\cdot) &= \frac{1}{N(\cdot)} \sum_{i \in N(\cdot)} \left[y_i - c(\cdot)\right]^2,
\end{align}
\end{subequations}
with $N(\cdot)$ representing the number of covariate-outcome pairs at the node of interest.

DTs are readily fitted using exhaustive search for each split. Cross-validation, usually on the maximal depth of the tree or the minimal number of samples per node, avoids overfitting of the tree. The optimized tree is then used for forecasting the outcome by taking the average of all outcomes belonging to a given node $m$. This yields a decision tree with piecewise constant predictions.

%
%


\section{Mixture Models}
\label{sec:Mixture_Models}
In this section, we describe the fitting procedure of CGMMs on data that combine multiple linear regression models to act as an ensemble learner. Given a set of covariate-outcome pairs (in our case $y_i$ denotes energy consumption),
\begin{equation}\mathcal{D} = \lbrace (x_i, y_i): i = 1, \ldots, N\rbrace,
\end{equation}
the idea is to model the probability distribution of any observation $y$ with corresponding covariates $x$ as the output of an ensemble of linear regressions
\begin{equation}\label{eq:mixture_distribution}
\mathbb{P}(y | x, \underbrace{\mathbf{w}, \sigma^2, \mathbf{\pi}}_{=:\theta}) = \sum_{k=1}^K \pi_k \mathcal{N}(y|w_k \cdot x, \sigma^2),
\end{equation}
where $\pi = \lbrace \pi_1, \ldots, \pi_K \rbrace$ and $\mathbf{w} = \lbrace w_1, \ldots, w_K \rbrace$ denote $K$ mixing proportions with $\sum_{i=1}^K \pi_k = 1$ and the regression coefficients for each learner, respectively. $\sigma^2$ signifies the noise variance, where, according to \cite{Bishop:2006aa}, we make the following
\begin{assumption}\label{as:sigma2_constant}
\textit{$\sigma^2$ is equal across all mixture components $k = 1, \ldots, K$.}
\end{assumption}
Assumption \ref{as:sigma2_constant} can be relaxed by using mixture-specific noise covariances $\lbrace \sigma_1^2, \ldots, \sigma_K^2 \rbrace$, in which case \eqref{eq:Mixture_update_1}$-$\eqref{eq:Mixture_update_4} need to be modified.

%
%


\subsection{Parameter Estimation}
Given the training data $\mathcal{D}$, the Expectation-Maximization Algorithm (EM-Algorithm) \cite{Rabiner:1989aa, Bishop:2006aa} allows us to derive an iterative procedure to learn the parameters $\theta = \lbrace\lbrace \pi_k \rbrace_{k=1}^K, \lbrace w_k \rbrace_{k=1}^K, \sigma^2 \rbrace$. We first define the expected complete log likelihood $\ell(\theta|\mathcal{D}_c)$, where 
\begin{equation}\mathcal{D}_c = \lbrace (x_i, y_i, z_i): i = 1, \ldots, N\rbrace
\end{equation}
denotes the fully observed dataset whose latent variables $\lbrace z_1, \ldots, z_N \rbrace$ are assumed to be known. The latent variable belonging to $x_i$ is a vector $z_i = \left[z_{i1}, \ldots, z_{iK}\right]^\top$, where $z_{ik}$ denotes the probability that $x_i$ was generated by mixture component $k$. The complete log-likelihood is
\begin{equation}\label{eq:mixture_log_likelihood}
\ell(\theta|\mathcal{D}_c) = \sum_{i=1}^N \sum_{k=1}^K z_{ik}\log\left(\pi_k\mathcal{N}(y_i|w_k \cdot x_i, \sigma^2)\right)
\end{equation}
under the assumption of known $z_{ik}$. The EM-Algorithm alternates between the E-Step, whose task is to determine the expected value of the latent variables $z_{ik}, 1\leq i\leq N, 1\leq k \leq K$ with respect to the conditional probability distribution \eqref{eq:mixture_distribution}, and the M-Step, which updates the parameters $\theta$ with the results from the E-Step by taking the derivative of the expected value of \eqref{eq:mixture_log_likelihood} with respect to the desired parameters $\theta$. This is carried out iteratively until some convergence criterion is reached, i.e. the incremental increase of the expected complete log likelihood \eqref{eq:mixture_log_likelihood} falls below a threshold. The update steps for one iteration are as follows:
\begin{subequations}
\begin{align}
\hat{z}_{ik} &= \frac{\hat{\pi}_k \mathcal{N}(y_i|\hat{w}_k \cdot x_i, \hat{\sigma}^2)}{\sum_{j=1}^K \hat{\pi}_j \mathcal{N}(y_i|\hat{w}_j \cdot x_i, \hat{\sigma}^2)}, \label{eq:Mixture_update_1}\\
\hat{\pi}_k &= \frac{1}{N} \sum_{i=1}^N \hat{z}_{ik}, \\
\hat{w}_k &= \left[ X^\top D X\right]^{-1}X^\top D Y,~ D = \text{diag}(\hat{z}_{1k}, \ldots, \hat{z}_{Nk}), \\
\hat{\sigma}^2 &= \frac{1}{N} \sum_{i=1}^N \sum_{k=1}^K \hat{z}_{ik} (y_i - \hat{w}_k\cdot x_i)^2, \label{eq:Mixture_update_4}
\end{align}
\end{subequations}
where we have to incorporate the constraint $\sum_{k=1}^K \hat{\pi}_k = 1$ as a Lagrange Multiplier in the derivation.

\subsection{Predicting New Data}

To predict the outcome $\hat{y}$ of an out-of-sample data point $x$, we suggest a different approach than is employed by \cite{Bishop:2006aa}: Instead of using the estimated mixing proportions $\lbrace \hat{\pi}_k \rbrace_{k=1}^K$ as the weights for a convex combination of the estimated regression coefficients $\lbrace \hat{w}_k \rbrace_{k=1}^K$, we choose the weights as the estimated latent variables $\lbrace \hat{z}_{jk}\rbrace_{k=1}^K$ of $x$'s nearest neighbor $x_j$:
\begin{subequations}
\begin{align}
j &= \arg\min_{1\leq i \leq N}\Vert x_i - x\Vert_2 \label{eq:CGMM_prediction_1}\\
\hat{y} &= \sum_{k=1}^K \hat{z}_{jk} \hat{w}_k \cdot x \label{eq:CGMM_prediction_2}
\end{align}
\end{subequations}
The rationale behind this approach is to exploit potential spatial separation in the set of training data, i.e. the fact that different regions of the covariate space are best fit by a specific learner. By locating the nearest neighbor of $x$, the same set of weights that proved to be most accurate for the training of the data points in the region around $x$ are to be used for the prediction of $\hat{y}$.


%
%


\section{Hidden Markov Models}
\label{sec:HMMs}
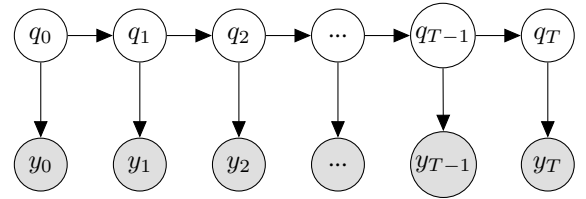
\begin{figure}[hbtp]
\vspace{-0.3cm}
\centering

\begin{tikzpicture}
\node[latent] 						(q0) {$q_0$};
\node[latent, right=of q0, xshift=-0.4cm] 	(q1) {$q_1$};
\node[latent, right=of q1, xshift=-0.4cm]	(q2) {$q_2$};
\node[latent, right=of q2, xshift=-0.4cm]	(dots) {...};
\node[latent, right=of dots, xshift=-0.4cm] 	(qt1) {$q_{T-1}$};
\node[latent, right=of qt1, xshift=-0.4cm]	(qt) {$q_{T}$};

\node[obs, below=of q0] 				(y0) {$y_0$};
\node[obs, below=of q1] 				(y1) {$y_1$};
\node[obs, below=of q2] 				(y2) {$y_2$};
\node[obs, below=of dots] 			(dts) {...};
\node[obs, right=of dts, xshift=-0.4cm]	(yt1) {$y_{T-1}$};
\node[obs, below=of qt] 				(yt) {$y_{T}$};

\edge {q0} {q1};
\edge {q1} {q2};
\edge {q2} {dots};
\edge {dots} {qt1};
\edge {qt1} {qt};
\edge {q0} {y0};
\edge {q1} {y1};
\edge {q2} {y2};
\edge {dots} {dts};
\edge {qt1} {yt1};
\edge {qt} {yt};

\end{tikzpicture}
\vspace{0.1cm}
\caption{Hidden Markov Model. Hidden States $q$, Observations $y$}
\vspace{-0.3cm}
\label{fig:HMM_schematic}
\end{figure}

In this section, we briefly outline the training procedure of HMMs. Figure \ref{fig:HMM_schematic} shows the graphical model of a standard HMM with a hidden layer (transparent nodes), representing latent variables, and observations (shaded nodes).

\subsection{Hidden Layer}\label{sec:Discrete_Markov_Processes}
We model the latent variables in the hidden layer (see Figure \ref{fig:HMM_schematic}) as a first order, time-invariant, Discrete Time Markov Chain (DTMC) with a set of transition probabilities
\begin{equation}
a_{ij} = \mathbb{P}(q_t=j|q_{t-1}=i),\quad~ 1\leq i,j \leq M,
\end{equation}
where $t=0, 1, 2, \ldots, T$ denote time instants associated with state changes and $q_t$ the hidden state at time $t$. Due to the Markov Property, we have that, conditional on $q_t$, $q_{t+1}$ is independent of $q_{t-1}$. The state transition coefficients $a_{ij}$ have the properties
\begin{equation}\label{eq:MC_transition_constraints}
0 \leq a_{ij}\leq 1, \quad \sum_{j=1}^M a_{ij}=1, \quad ~i,j \in \lbrace 1, \ldots, M\rbrace,
\end{equation}
where $M$ denotes the number of states (=latent variables).

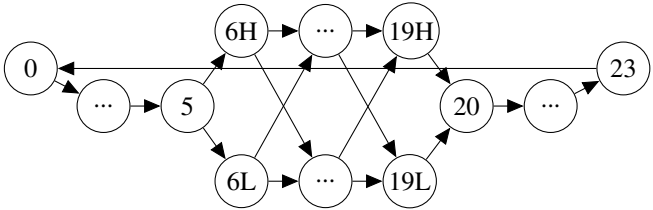
\begin{figure}[hbtp]
\vspace{-0.2cm}
\centering
\begin{tikzpicture}
\node[latent] (1) {0};
\node[latent, right=of 1, xshift=-0.75cm, yshift=-0.5cm]	(dots) {...};
\node[latent, right=of dots, xshift=-0.6cm]	(5) {5};
\node[latent, right=of 5, yshift=1cm, xshift=-1.0cm] (6H) {6H};
\node[latent, right=of 5, yshift=-1cm, xshift=-1.0cm] (6L) {6L};
\node[latent, right=of 6H, xshift=-0.6cm] (2dots) {...};
\node[latent, right=of 6L, xshift=-0.6cm] (3dots) {...};
\node[latent, right=of 2dots, xshift=-0.6cm] (19H) {19H};
\node[latent, right=of 3dots, xshift=-0.6cm] (19L) {19L};
\node[latent, right=of 19H, yshift=-1cm, xshift=-1.0cm] (20) {20};
\node[latent, right=of 20, xshift=-0.6cm] (4dots) {...};
\node[latent, right=of 4dots, xshift=-0.75cm, yshift=0.5cm]	(23) {23};

\edge {} {1};
\edge {1} {dots};
\edge {dots} {5};
\edge {5} {6H, 6L};
\edge {6H} {2dots, 3dots};
\edge {6L} {2dots, 3dots};
\edge {2dots, 3dots} {19H};
\edge {2dots, 3dots} {19L};
\edge {19H, 19L} {20};
\edge {20} {4dots};
\edge {4dots} {23};
\edge {23} {1}

\end{tikzpicture}
\vspace{0.01cm}
\caption{Markov State Transition Diagram, 24 Hour Periodicity. For Example, ``5" Signifies Time Between 5 a.m. - 6 a.m.}
\vspace{-0.3cm}
\label{fig:MC_transition_diagram}
\end{figure}

We postulate the existence of two different latent states for each hour of the day (HoD) between 6 a.m. - 8 p.m., and a single state for the remaining hours, hence $M = 38$. For the former hours, binary states describing each hour shall encode information about ``high" (``H") or ``low" (``L") consumption, which might be an indicator for occupancy (``H" = at home, ``L" = not at home). For the remaining HoDs, we note that first, no DR events in our data set were recorded outside this window, and second, little variation in the smart meter recordings was observed, which is consistent with \cite{Zhou:2016aa}, where the authors find little variation in clustered load shapes during the night. Due to the Markov Property, state transitions are restricted to states belonging to the next hour only, which renders the Markov transition matrix $A \in \mathbb{R}^{38\times 38}$ sparse. Figure \ref{fig:MC_transition_diagram} shows the state transition diagram (without probabilities on the edges, which are to be estimated from data, see Section \ref{sec:Parameter_Estimation}).

A logical extension is to allow for multi-step dependencies, which can be achieved by enlarging the state space of the DTMC such that the previous $n>1$ states jointly determine the next transition. A more granular description of the state transitions, however, would come at the cost of a higher computational complexity, a tradeoff whose analysis is outside the scope of this paper.

A consequence of this modeling approach is that, if the consumption is high at time $t-1$, it is likely that the hidden state $q_{t-1} = \text{H}$ and $q_t=\text{H}$, and so we expect a high consumption at time $t$, as well. Conversely, if the consumption at time $t-1$ is low (i.e. due to an absent user), the most likely hidden state $q_{t-1}=\text{L}$ and $q_t=\text{L}$, and thus we would expect a low consumption at time $t$. It turns out that the parameter estimation on the data set used in Section \ref{sec:Real_Data} automatically assigns higher probabilities for transitions to the next hour of the same type than to the opposite type, indicating that switches between ``H" and ``L" do not occur frequently. This is consistent with our intuition: If the latent variable represents periods of expected presence or absence at home, users are more likely to remain either at home or absent, rather than switching every hour.

\subsection{Observations}
\begin{assumption}\label{as:emission_HMM}
\textit{Conditional on the current hidden state $q_t$, the observable energy consumption $y_t$ (=observation/emission) is assumed to be normally distributed with parameters $(\mu_{q_t}, \sigma_{q_t}^2)$}:
\begin{equation}\label{eq:emission_distribution}
\mathbb{P}(y_t | q_t) = \frac{1}{\sqrt{2\pi\sigma_{q_t}^2}}\exp\left(-\frac{(y_t-\mu_{q_t})^2}{2\sigma_{q_t}^2}\right).
\end{equation}
\end{assumption}
An obvious extension is to choose alternative distributions, an idea we do not investigate further in this paper.

\subsection{Parameter Estimation and Inference}\label{sec:Parameter_Estimation}
Given an observed sequence of emissions $Y := \lbrace y_0, y_1, \ldots, y_T \rbrace$ with known initial state distribution $\pi_{q_0}$, the parameters of the HMM $\theta := \lbrace\lbrace a_{ij}\rbrace, \lbrace\mu_{q_t}\rbrace, \lbrace\sigma_{q_t}^2 \rbrace \rbrace$, i.e. the transition probabilities and emission parameters, can be estimated with the EM-Algorithm. Starting from the complete log-likelihood
\begin{equation}\label{eq:log_likelihood}
\ell(\theta | \mathcal{D}_c) = \log\left( \pi_{q_0} \prod_{t=0}^{T-1}a_{q_t,q_{t+1}} \prod_{t=0}^T \mathcal{N}(y_t | \mu_{q_t}, \sigma_{q_t}^2)\right),
\end{equation}
with the fully observed data set
\begin{equation}
\mathcal{D}_c = \lbrace (y_n, q_n, a_{q_n,q_{n+1}}) : n \in \left[0, T-1\right] \rbrace \cup \lbrace \pi_{q_0}, y_T, q_T \rbrace,
\end{equation}
minimizing the expected value of \eqref{eq:log_likelihood} with respect to the desired variables $\theta$ to be estimated yields the update equations for the M-Step of the EM-algorithm (also called Baum-Welch Updates):
\begin{subequations}
\begin{align}
\hat{\pi}_i &= \mathbb{P}(q_0=i|Y) \label{eq:pi_0_update}\\
\hat{a}_{ij} &= \frac{\sum_{t=0}^{T-1}\mathbb{P}(q_t=i,q_{t+1}=j|Y)}{\sum_{t=0}^{T-1}\sum_{j=1}^M \mathbb{P}(q_t=i,q_{t+1}=j|Y)} \label{eq:a_ij_update}\\
\hat{\mu}_i &= \frac{\sum_{t=0}^T y_t\cdot\mathbb{P}(q_t=i|Y)}{\sum_{t=0}^T\mathbb{P}(q_t=i|Y)} \label{eq:mu_update}\\
\hat{\sigma}_i^2 &= \frac{\sum_{t=0}^T \mathbb{P}(q_t=i|Y) (y_t - \hat{\mu}_i)^2}{\sum_{t=0}^T \mathbb{P}(q_t=i|Y)} \label{eq:sigma_update}
\end{align}
\end{subequations}
To arrive at Equations \eqref{eq:pi_0_update} and \eqref{eq:a_ij_update}, the stochastic constraints described in \eqref{eq:MC_transition_constraints} and sparsity patterns of the transition matrix $A$ as well as $\sum_{i=1}^M \pi_i = 1$ are used as Lagrange multipliers during the minimization of  
\eqref{eq:log_likelihood}.

Using Bayes Rule, the E-Step of the EM-algorithm computes the sufficient statistics $\mathbb{P}(q_t=i,q_{t+1}=j|Y)$ and $\mathbb{P}(q_t=i|Y)$ with the well-known \textit{Alpha-Beta-Recursion}:
\begin{align}
\mathbb{P}(q_t|Y) &= \frac{\mathbb{P}(Y|q_t)\mathbb{P}(q_t)}{\mathbb{P}(Y)} \nonumber\\
&=\frac{\mathbb{P}(y_0,\ldots,y_{t-1},q_t)\mathbb{P}(y_t|q_t)\mathbb{P}(y_{t+1},\ldots,y_T|q_t)}{\mathbb{P}(Y)}\nonumber\\
&=:\frac{\alpha(q_t)\mathbb{P}(y_t|q_t)\beta(q_t)}{\mathbb{P}(Y)}. \label{eq:P(q_t|y)}
\end{align}
We note that $\alpha(q_t)$ is defined as $\mathbb{P}(y_0,\ldots,y_{t-1},q_t)$ rather than $\mathbb{P}(y_0,\ldots,y_t,q_t)$ as is done in \cite{Rabiner:1989aa, Jordan:2007aa}. This is done for a simplified treatment of its update step \eqref{eq:alpha_update} and the prediction problem \eqref{eq:prediction}.

Using Bayes Rule, $\alpha(q_t)$ and $\beta(q_t)$ can be updated recursively:
\begin{align}
\alpha(q_{t+1}) &= \mathbb{P}(y_0, \ldots, y_{t},q_{t+1}) \nonumber\\ 
&=\sum_{q_t}\mathbb{P}(y_0, \ldots, y_{t},q_t,q_{t+1}) \nonumber\\
&=\sum_{q_t}\mathbb{P}(y_0, \ldots, y_{t-1}|q_t)\mathbb{P}(y_t|q_t)\mathbb{P}(q_{t+1}|q_t)\nonumber\\
&=\sum_{q_t}\alpha(q_t)\mathbb{P}(y_t|q_t)a_{q_t,q_{t+1}}. \label{eq:alpha_update}\\
\beta(q_{t}) &= \mathbb{P}(y_{t+1},\ldots,y_T|q_t) \nonumber\\
&=\sum_{q_{t+1}}\mathbb{P}(y_{t+1},\ldots,y_T,q_{t+1}|q_t)\nonumber\\
&=\sum_{q_{t+1}}\mathbb{P}(y_{t+2},\ldots,y_T|q_{t+1})\mathbb{P}(y_{t+1}|q_{t+1})\mathbb{P}(q_{t+1}|q_t)\nonumber\\
&=\sum_{q_{t+1}}\beta(q_{t+1})\mathbb{P}(y_{t+1}|q_{t+1})a_{q_t,q_{t+1}}. \label{eq:beta_update}
\end{align}
Note that $\mathbb{P}(y_t|q_t),~ 0 \leq t \leq T$ can be computed with \eqref{eq:emission_distribution}.

$\alpha(q_1)$ is initialized as $\pi_{q_0}$ and $\beta(q_T)$ as a vector of ones.

With the definition of $\alpha(q_t)$ and $\beta(q_t)$, $\mathbb{P}(q_t,q_{t+1}|Y)$ is computed as follows:
\begin{align}
&\mathbb{P}(q_t,q_{t+1}|Y) = \frac{\mathbb{P}(Y|q_t,q_{t+1})\mathbb{P}(q_t,q_{t+1})}{\mathbb{P}(Y)} \nonumber \\
&= \frac{\mathbb{P}(y_0, \ldots, y_{t-1}|q_t)\mathbb{P}(y_t|q_t)\mathbb{P}(y_{t+1}|q_{t+1})}{\mathbb{P}(Y)}\times \nonumber \\
&\quad~\mathbb{P}(y_{t+2}, \ldots, y_T|q_{t+1})\mathbb{P}(q_t,q_{t+1})\nonumber\\
&= \frac{\alpha(q_t)\beta(q_{t+1})a_{q_t,q_{t+1}}\mathbb{P}(y_t|q_t)\mathbb{P}(y_{t+1}|q_{t+1})}{\mathbb{P}(Y)}. \label{eq:P(q_t,q_t+1|q)}
\end{align}

In summary, the EM-algorithm iterates between the E-Step to compute the sufficient statistics $\mathbb{P}(q_t=i,q_{t+1}=j|Y)$ and $\mathbb{P}(q_t=i|Y)$ with Equations \eqref{eq:P(q_t|y)}, \eqref{eq:alpha_update}, \eqref{eq:beta_update}, and \eqref{eq:P(q_t,q_t+1|q)} while fixing the parameters in \eqref{eq:pi_0_update}$-$\eqref{eq:sigma_update}, and the M-Step to update the parameters in \eqref{eq:pi_0_update}$-$\eqref{eq:sigma_update} while fixing the sufficient statistics  until some convergence criterion on the expected value of \eqref{eq:log_likelihood} is reached.

\subsection{Filtering, Smoothing, and Predicting the Latent Variable}\label{sec:Filtering_Smoothing_Predicting}
After the parameters of the HMM have been estimated, we turn to the problem of estimating the probabilities of the most likely hidden state. Given the observation sequence $Y := \lbrace y_0, y_1, \ldots, y_T \rbrace$, the \textit{filtering problem} calculates $\mathbb{P}(q_T|Y)$:
\begin{align}
\mathbb{P}(q_T|y_0, \ldots, y_T) &= \frac{\mathbb{P}(y_0, \ldots, y_T|q_T)\mathbb{P}(q_T)}{\mathbb{P}(y_0, \ldots, y_T)} \nonumber \\
&= \frac{\mathbb{P}(y_0, \ldots, y_{T-1}| q_T)\mathbb{P}(y_T|q_T)\mathbb{P}(q_T)}{\mathbb{P}(y_0, \ldots, y_{T})}\nonumber \\
&= \frac{\alpha(q_T)\mathbb{P}(y_T|q_T)}{\mathbb{P}(y_0, \ldots, y_T)}.\label{eq:filtering}
\end{align}

Alternatively, the \textit{prediction problem} can be used to predict the probability of the next hidden state at time $T+1$, i.e.
\begin{align}
\mathbb{P}(q_{T+1}|y_0, \ldots, y_T) &= \frac{\mathbb{P}(y_0, \ldots, y_T|q_{T+1})\mathbb{P}(q_{T+1})}{\mathbb{P}(y_0, \ldots, y_T)} \nonumber \\
&= \frac{\alpha(q_{T+1})}{\mathbb{P}(y_0, \ldots, y_T)}. \label{eq:prediction}
\end{align}

Lastly, the \textit{smoothing problem} can be solved to ex-post predict the probability of the latent variable at a past time $1 \leq p < T$:
\begin{align}
&\mathbb{P}(q_p|y_0, \ldots, y_T) = \frac{\mathbb{P}(y_0, \ldots, y_T|q_p)\mathbb{P}(q_p)}{\mathbb{P}(y_0, \ldots, y_T)} \nonumber \\
&= \frac{\mathbb{P}(y_0, \ldots, y_{p-1}| q_p)\mathbb{P}(y_p|q_p)\mathbb{P}(y_{p+1}, \ldots, y_T|q_p)\mathbb{P}(q_p)}{\mathbb{P}(y_0, \ldots, y_{T})}\nonumber \\
&= \frac{\alpha(q_p)\mathbb{P}(y_p|q_p)\beta(q_p)}{\mathbb{P}(y_0, \ldots, y_T)}.\label{eq:smoothing}
\end{align}

%
%


\section{Short-Term Load Forecasting}
\label{sec:STLF}
In the following, we describe \textit{online} forecasting algorithms that allow for including knowledge about the estimated latent variables obtained from HMMs and CGMMs into the ML methods introduced in Section \ref{sec:Forecasting_Methods}. We make the following
\begin{assumption}\label{as:stationarity}
\textit{The consumption time series $Y$ is} stationary, \textit{i.e. there are no structural changes in consumption behavior over time.}
\end{assumption}
This assumption is sound as we explain in Section \ref{sec:Real_Data}.

\subsection{Covariates for Prediction}\label{sec:Covariates_for_Prediction}
The following observable covariates are used for all forecasting methods: 
\begin{itemize}
\item Five previous hourly consumptions
\item Five previous hourly ambient air temperatures
\item A categorical variable for the hour of day for ML methods without latent variable and the CGMM
\item A categorical variable interacting the hour of day with the estimated latent variable obtained from HMM for ML methods with HMM
\end{itemize}

\subsection{Prediction with Hidden Markov Model}\label{sec:HMM_Prediction_Procedure}
%
%

\begin{algorithm}
 \caption{Algorithm for Online Prediction with HMM}
 \begin{algorithmic}[1]\label{alg:RT_prediction}
 \renewcommand{\algorithmicrequire}{\textbf{Input:}}
 \renewcommand{\algorithmicensure}{\textbf{Output:}}
 \REQUIRE Training Data $\mathcal{D}_{\text{tr}}:= \lbrace (x_t, y_t) : t = 0, \ldots, T \rbrace$, Test Data $\mathcal{D}_{\text{te}}:= \lbrace (x_t, y_t) : t = T+1, \ldots, \tau \rbrace$, ML Method
 \STATE Initialize all $\mu_1, \ldots, \mu_{38}, \sigma_1^2, \dots, \sigma_{38}^2$ suitably
 \STATE Initialize all $a_{ij}$, observing \eqref{eq:MC_transition_constraints} and Figure \ref{fig:MC_transition_diagram}
 \WHILE {$\Delta \mathbb{E}\left[\ell(\theta | \mathcal{D}_c) \right] < \epsilon$}
 \STATE Do E-Step: Calculate \eqref{eq:emission_distribution} and \eqref{eq:P(q_t,q_t+1|q)} for $t = \left[0, \ldots, T-1\right],\ q_t, q_{t+1} = \left[1, \ldots, 38\right]$ with \eqref{eq:P(q_t|y)}$-$\eqref{eq:beta_update}
 \STATE Do M-Step: Update HMM parameters with \eqref{eq:pi_0_update}$-$\eqref{eq:sigma_update}
 \ENDWHILE
 \STATE Solve smoothing problem \eqref{eq:smoothing} for $t=0, \ldots, T-1$ 
 \STATE Solve filtering problem \eqref{eq:filtering} for $t=T$ 
 \STATE Round $\mathbb{P}(\hat{q}_0 | \mathcal{D}_{\text{tr}}), \ldots, \mathbb{P}(\hat{q}_T | \mathcal{D}_{\text{tr}})$ to 0 / 1
 \STATE Fit ML Method on $\lbrace ((x_t, \mathbb{P}(\hat{q}_t)), y_t) : t \in 0, \ldots, T \rbrace $
  
 \FOR {$s$ in $[T+1, \tau]$}
 \STATE Solve prediction problem \eqref{eq:prediction} at time $s$
 \STATE Round $\mathbb{P}(\hat{q}_s)$ to 0 / 1
 \STATE Predict $\hat{y}_s$ with ML method on covariates $(x_s,\mathbb{P}(\hat{q}_s))$
 \ENDFOR

 \RETURN $\hat{y}_{T+1}, \ldots, \hat{y}_\tau$ 
 \end{algorithmic} 
 \end{algorithm}

Algorithm \ref{alg:RT_prediction} describes the procedure of fitting an HMM on training data $\mathcal{D}_{\text{tr}}$, which yields estimated latent variables to be used as additional covariates for the ML methods presented in Section \ref{sec:Forecasting_Methods} to perform stepwise prediction on the covariates of the test data $\mathcal{D}_{\text{te}}$. The prediction accuracy of these outcomes is then compared to those outcomes predicted by ML methods that are trained on the training data $\mathcal{D}_{\text{tr}}$ without estimated latent variables in the covariates.

\subsection{Prediction with Conditional Gaussian Mixture Model}

Algorithm \ref{alg:RT_prediction_CGMM} describes the online prediction method for a CGMM with $k=2$ on a given set of training and test data. $\hat{w}$ obtained by OLS is perturbed with zero mean Gaussian Noise $\epsilon$ to obtain the initializations $w_1, w_2$. Note that this step is necessary to break the symmetry of the update steps \eqref{eq:Mixture_update_1}$-$\eqref{eq:Mixture_update_4}, which would keep $w_1 = w_2 = \hat{w}$ unchanged.

\begin{algorithm}
 \caption{Algorithm for Online Prediction with CGMM}
 \begin{algorithmic}[1]\label{alg:RT_prediction_CGMM}
 \renewcommand{\algorithmicrequire}{\textbf{Input:}}
 \renewcommand{\algorithmicensure}{\textbf{Output:}}
 \REQUIRE Training Data $\mathcal{D}_{\text{tr}}:= \lbrace (x_t, y_t) : t = 0, \ldots, T \rbrace$, Test Data $\mathcal{D}_{\text{te}}:= \lbrace (x_t, y_t) : t = T+1, \ldots, \tau \rbrace$
 \STATE Fit OLS model \eqref{eq:X_y_linear} on $\mathcal{D}_{\text{tr}}$ to obtain $\hat{w}$
 \STATE Initialize $w_1 \leftarrow \hat{w} + \epsilon$
 \STATE Initialize $w_2 \leftarrow \hat{w} + \epsilon$
 \WHILE {$\Delta \mathbb{E}\left[ \ell(\theta | \mathcal{D}_c)\right] < \epsilon$}
 \STATE Update CGMM parameters \eqref{eq:Mixture_update_1}$-$\eqref{eq:Mixture_update_4}
 \ENDWHILE

 \FOR {$s$ in $[T+1, \tau]$}
 \STATE Predict $\hat{y}_s$ with \eqref{eq:CGMM_prediction_1} and \eqref{eq:CGMM_prediction_2}
 \ENDFOR
 \RETURN $\hat{y}_{T+1}, \ldots, \hat{y}_\tau$ 
 \end{algorithmic} 
 \end{algorithm}
 
Note that in both Algorithms \ref{alg:RT_prediction} and \ref{alg:RT_prediction_CGMM}, the model-specific parameters could be updated after each prediction as more data from the test sequence is observed and hence enters $\mathcal{D}_{\text{tr}}$.

\subsection{Metric for Forecasting Accuracy}
The Mean Absolute Percentage Error (MAPE) of predictions of a set of discrete values $v_i \in \mathcal{V}$ is used to evaluate the accuracy of the predictor:
\begin{equation}\label{eq:MAPE}
\mathrm{MAPE} = \frac{1}{|\mathcal{V}|}\sum_{i\in \mathcal{V}} \left| \frac{\hat{v}_i-v_i}{v_i}\right|\cdot 100 \%,
\end{equation}
where $\hat{v}_i$ denotes the estimate of $v_i$.

\section{Non-Experimental Estimates of DR Reduction}
\label{sec:Reductions}
To estimate individual treatment effects, we adopt the potential outcomes framework \cite{Rubin:1974aa} with binary treatments $T_t \in \{0,1\}$, where $T_t = 1$ corresponds to prompting the user to reduce consumption at time $t$, and $T_t = 0$ denotes the absence of any design intervention, hence ``control". Let $y_t^0$ and $y_t^1$ denote the response (i.e. the electricity consumption) that would be observed if an individual received treatment 0 and 1 at time $t$, respectively. The goal is to estimate the conditional treatment effect, i.e. 
\begin{equation}\label{eq:Conditional_Treatment_Effect}
\Delta(x) = \mathbb{E}\left[y^1 | x \in \mathcal{X}\right] - \mathbb{E} \left[y^0 | x \in \mathcal{X}\right],
\end{equation}
where $x$ denotes a vector of observable covariates in the covariate space $\mathcal{X}$. Assuming an unconfounded assignment mechanism of treatments to individuals and independency of the potential outcomes of the time index $t$, conditional on the covariates (see \cite{Rubin:1974aa} for details), the true causal effect of DR, namely $\left(y_t^0 - y_t^1\right)$, cannot be found because only one of $y_t^0$ and $y_t^1$ can be observed (c.f. Fundamental Problem of Causal Inference \cite{Holland:1986aa}).

Causal Inference can thus be interpreted as a ``Missing Data Problem". Given the observed treatment outcomes $y_{t_1}^1, \ldots, y_{t_n}^1$ (i.e. observed consumptions during DR hours $t_1, \ldots, t_n$), to estimate the true causal effect of treatment, one would require a credible estimate of the \textit{counterfactuals} $\hat{y}_{t_1}^0, \ldots, \hat{y}_{t_n}^0$, i.e. the outcome in the hypothetical absence of treatment, to be able to compute the conditional treatment effect \eqref{eq:Conditional_Treatment_Effect}.

To compute such estimates in a non-experimental way, we split the available consumption data into a pretreatment training set with time indices $t \in \mathcal{P}$ consisting of ``regular" electricity consumption, i.e. all smart meter readings before the customers' signup date with the DR provider, and a test set with corresponding times $t \in \mathcal{S}$ thereafter which itself consists of smart meter readings during DR hours $\mathcal{T}$ (treatment) and outside DR hours $\mathcal{C}$ (control), hence $\mathcal{S} = \mathcal{T} \cup \mathcal{C}$. Let
\begin{subequations}
\begin{align}
\mathcal{D}_P &= \lbrace\left(x_{i,t}^0, y_{i,t}^0\right) : t \in \mathcal{P}\rbrace \label{eq:pretreatment_data}\\
\mathcal{D}_C &= \lbrace\left(x_{i,t}^0, y_{i,t}^0\right) : t \in \mathcal{C}\rbrace \label{eq:control_data}\\
\mathcal{D}_T &= \lbrace\left(x_{i,t}^1, y_{i,t}^1\right) : t \in \mathcal{T}\rbrace \label{eq:treatment_data}
\end{align}
\end{subequations}
denote covariate/outcome pairs for the pretreatment period, the control observations, and the treatment observations of user $i$, respectively. Note that the set of treatment outcomes $\lbrace y_{i,t}^1 : t \in \mathcal{T}\rbrace$ captures the hourly electricity consumption of user $i$ during DR events and is likely to deviate from the ``usual" consumption, i.e. the user's consumption, had there been no treatment. By fitting any regression model presented in Section \ref{sec:Forecasting_Methods} on the pretreatment training data $\mathcal{D}_P$ of a given user $i$, and under Assumption \ref{as:stationarity}, applying this model on the treatment covariates $\lbrace x_{i,t}^1 : t \in \mathcal{T}\rbrace$ yields the estimated counterfactual consumptions $\lbrace\hat{y}_{i,t}^0 : t \in \mathcal{T}\rbrace$ for user $i$. In particular, Assumption \ref{as:stationarity} states that DR treatments are interpreted as transitory shocks that do not result in a change in the consumption behavior for $t \in \mathcal{C}$. An elementwise comparison of $\lbrace\hat{y}_{i,t}^0 : t \in \mathcal{T}\rbrace$ and $\lbrace y_{i,t}^1 : t \in \mathcal{T}\rbrace$ yields the pointwise estimated reduction of user $i$'s electricity consumption $\lbrace\hat{y}_{i,t}^\Delta : t \in \mathcal{T}\rbrace$ during each DR event:
\begin{equation}\label{eq:estimated_reduction}
\lbrace\hat{y}_{i,t}^\Delta : t \in \mathcal{T}\rbrace = \lbrace (\hat{y}_{i,t}^0 - y_{i,t}^1) : t \in \mathcal{T}\rbrace.
\end{equation}
$\hat{y}_{i,t}^\Delta > 0$ corresponds to an estimated reduction of $\hat{y}_{i,t}^\Delta$, and conversely, $\hat{y}_{i,t}^\Delta < 0$ implies an estimated increase by $\vert\hat{y}_{i,t}^\Delta\vert$.


%
%


\section{Experiments on Data}
\label{sec:Real_Data}
We conduct a case study on a data set of a residential DR program including residential customers in the western United States, collected between 2012 and 2014. Aligned with those readings are timestamps of notifications sent by the DR provider to the users that prompt them to reduce their consumption for a short period, typically until the next full hour. A subset of the users have smart home devices that can be remotely shut off by the DR provider with the users' consent. Ambient air temperature measurements were logged from publicly available data sources to capture the correlation between temperature and electricity consumption.

\subsection{Characteristics of Data and Data Preprocessing}
\label{sec:Data_Characteristics_Preprocessing}
Users with the following characteristics are excluded from the analysis:
\begin{itemize}
\item Users with residential solar photovoltaics (PV) 
\item Users with corrupt smart meter readings, i.e. unrealistically high recordings
\end{itemize}
The consumption series of the remaining users are then aligned with available temperature readings and mapped to the range $[0, 1]$ to be able to compare users on a relative level. The temperature data is standardized to zero mean and unit variance. Lastly, the pretreatment data is tested for stationarity with the augmented Dickey-Fuller Test \cite{Fuller:1995aa} to assert, with a significance level of more than 99\%, the absence of a unit root, which motivates Assumption \ref{as:stationarity}.

\subsection{Experiments on Semi-Synthetic Data}
\label{sec:Synthetic_Data}
As only one of $\lbrace y_{i,t}^0, y_{i,t}^1 \rbrace$ for a given user $i$ at time $t$ can be observed, we construct semisynthetic data for which both values and hence the true causal effect $(y_{i,t}^0 - y_{i,t}^1)$ are known. This allows us to evaluate the accuracy of predicted counterfactual consumptions and the ensuing non-experimental estimates of DR reduction \eqref{eq:estimated_reduction}. For this purpose, we take actual pretreatment training data $\mathcal{D}_P$ \eqref{eq:pretreatment_data} for each user $i$, which is free of any DR messages. Next, we split this training set into two pieces by introducing an artificial signup date $\tilde{t}$ valid across all users. We thus obtain a synthetic training set $\mathcal{\tilde{D}}_P = \lbrace\left(x_{i,t}^0, y_{i,t}^0\right) : t \in \mathcal{P}, t < \tilde{t}\rbrace$ and a synthetic test set $\mathcal{\tilde{D}}_S = \lbrace\left( x_{i,t}^0, y_{i,t}^0\right) : t \in \mathcal{P}, t\geq \tilde{t}\rbrace$ for user $i$. Next, a random subset $\mathcal{\tilde{T}}$ of all available time indices in the synthetic test set $\mathcal{\tilde{D}}_S$ between 6 a.m. - 8 p.m. is assigned a synthetic treatment, for which the respective consumption is decreased by a random value $ \in [0, \bar{c}]$. By doing so, both the treatment and control outcomes for $t \in \tilde{\mathcal{T}}$ become available, and so we obtain the semisynthetic data set
\begin{equation}\label{eq:semisynthetic_data}
\mathcal{\tilde{D}}_{\tilde{T}} := \lbrace\left(x_{i,t}^0, y_{i,t}^0, y_{i,t}^1 \right) : t \in \mathcal{\tilde{T}}\rbrace.
\end{equation}
Thus, any non-experimental estimate of the DR treatment effect for $t \in \mathcal{\tilde{T}}$ can be benchmarked on the known (synthetic) counterfactual $\lbrace y_{i,t}^0 : t \in \mathcal{\tilde{T}}\rbrace$.

This semisynthetic data set is used for two purposes. First, we evaluate the MAPE \eqref{eq:MAPE} of the estimators from Section \ref{sec:Forecasting_Methods}, with and without latent variables. This is done by training them on $\mathcal{D}_{\text{tr}}=\tilde{D}_P$ and testing on $\mathcal{D}_{\text{te}}=\tilde{D}_S$, which yields out-of-sample counterfactual consumptions $\lbrace\hat{y}_{i,t}^0 : t\in \mathcal{\tilde{T}}\rbrace$ across all users $i$, see Algorithms \ref{alg:RT_prediction} and \ref{alg:RT_prediction_CGMM}. Second, we conduct a comparison of the eventwise errors of estimated DR reductions for all ML methods with the HMM latent variable (CGMM is not considered further), which, for a given user $i$ at time $t$, are obtained as follows:
\begin{equation}\label{eq:DR_reds_computation}
\hat{y}_{i,t}^\Delta - y_{i,t}^\Delta = \left(\hat{y}_{i,t}^0 - y_{i,t}^1\right) - \left(y_{i,t}^0 - y_{i,t}^1\right) = \hat{y}_{i,t}^0 - y_{i,t}^0.
\end{equation}
The ground truth counterfactual $y_{i,t}^0$ is available for the semisynthetic data \eqref{eq:semisynthetic_data} by construction, but would be unavailable for real-world data.

\begin{figure}[hbtp]
\centering
\vspace{-0.25cm}
\includegraphics[scale=0.34]{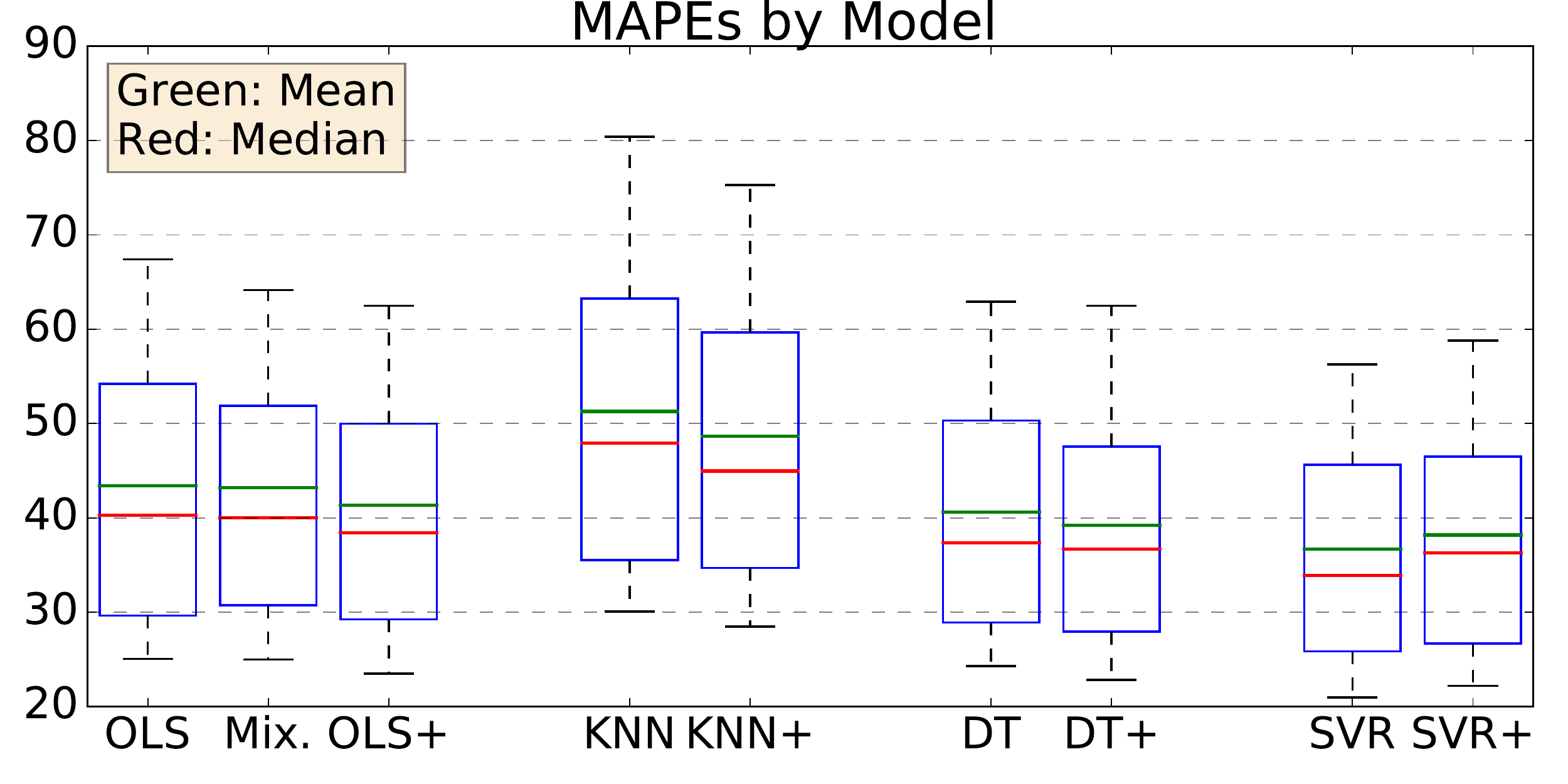}
\caption{Prediction Accuracy by Forecasting Method. ``+" Signifies Model with HMM Latent Variable, ``Mix." Denotes CGMM. Blue Boxes Span 25-75th Percentile, Whiskers 10-90th.}
\label{fig:MAPEs_Methods_Boxplot}
\end{figure}

Figure \ref{fig:MAPEs_Methods_Boxplot} shows a boxplot of the distribution of average MAPEs across users for the prediction methods introduced in Section \ref{sec:Forecasting_Methods} with and without the latent variable from HMM, and for the CGMM (Section \ref{sec:Mixture_Models}).
It can be seen that the information about the latent variable improves the prediction accuracy in all cases but SVR. Further, the lower MAPE obtained with DT and SVR is consistent with the findings in \cite{Mirowski:2014aa, Taylor:2007aa}. The higher MAPE for KNN compared to OLS can be explained by the different magnitudes of the covariates introduced in Section \ref{sec:Covariates_for_Prediction}, which gives categorical variables disproportionate weight. The CGMM performs better than OLS, but worse than OLS with the latent variable. Note that other more sophisticated predictors (e.g. Neural Networks) have lower MAPEs at the cost of longer computation times and potential loss of interpretability, but are likely to show a similar improvement in terms of MAPE by incorporating information about the estimated latent variable as the amount of training data increases. For a comparison between the prediction accuracy of state-of-the-art estimators, the reader is referred to \cite{Mirowski:2014aa, Taylor:2007aa} for further information.

\begin{figure}[hbtp]
\centering
\vspace{-0.25cm}
\includegraphics[scale=0.226]{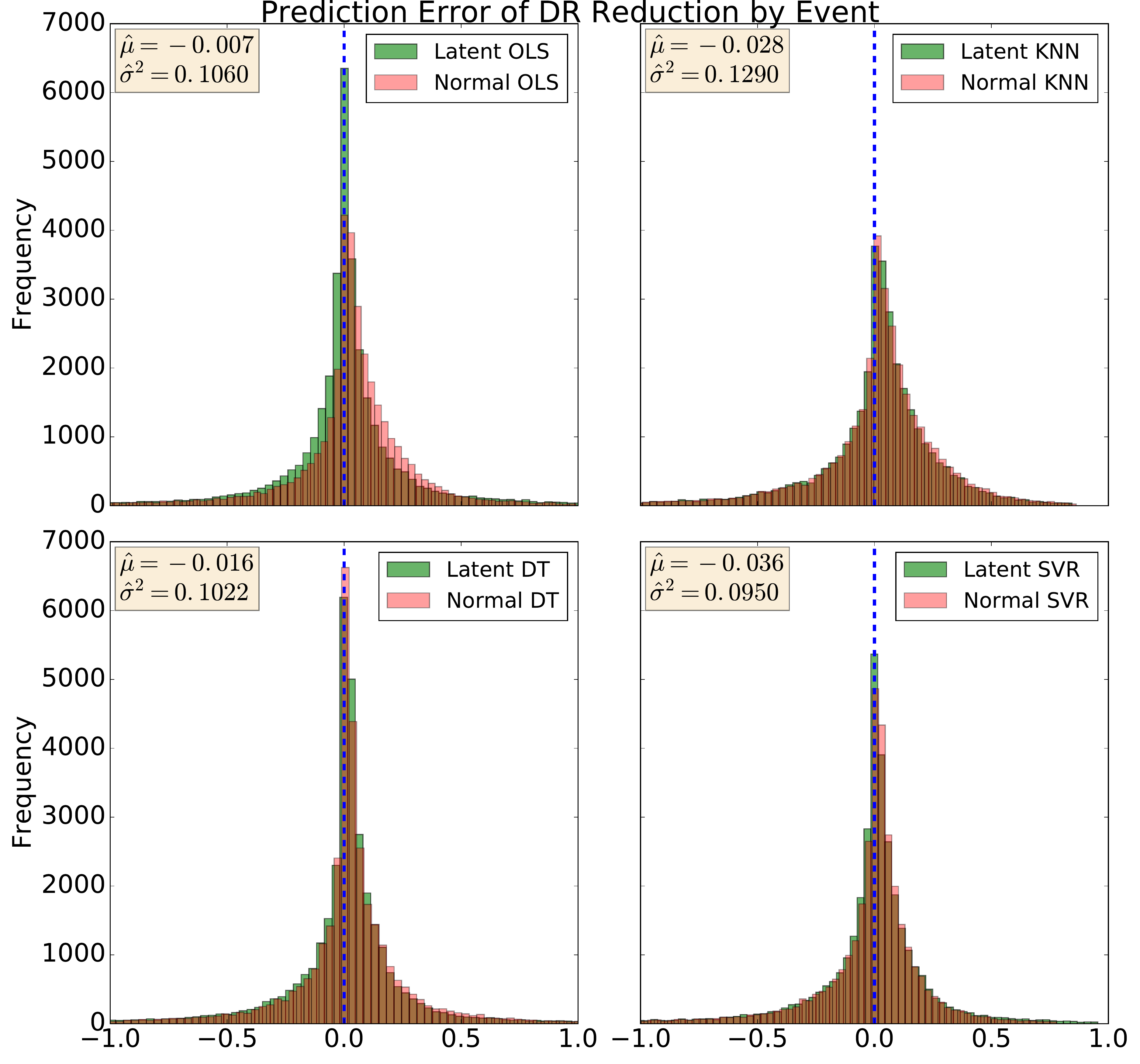}
\caption{Pointwise Prediction Error of DR Treatment Effect on the User Level; Bias $\hat{\mu}$ and Variance $\hat{\sigma}^2$ of Model with Latent Variable from HMM.}
\label{fig:Synthetic_Prediction_Errors}
\end{figure}

Figure \ref{fig:Synthetic_Prediction_Errors} shows histograms of eventwise prediction errors \eqref{eq:DR_reds_computation} in the estimated DR reduction for single events and across all users $i$. Green bars and red bars signify prediction errors of forecasting methods that do and do not make use of the estimated latent variable from HMM, respectively. Aligned with these plots are the sample mean and covariance of the errors for the models that take the latent variable into account. The bias-variance decomposition
\begin{equation}\label{eq:Bias_Variance_Tradeoff}
\mathbb{E}\left[ \left(\hat{y}_{i,t}^\Delta - y_{i,t}^\Delta\right)^2 \right] = \text{Bias}(\hat{y}_{i,t}^\Delta)^2 + \text{Var}(\hat{y}_{i,t}^\Delta) + \epsilon,
\end{equation}
where $\epsilon$ denotes the irreducible error, is invoked in the following. Noting that $\hat{\mu}$ and $\hat{\sigma}^2$ in Figure \ref{fig:Synthetic_Prediction_Errors} correspond to the bias and variance in \eqref{eq:Bias_Variance_Tradeoff} from the model with latent variable from HMM, the tradeoff becomes clear when comparing OLS, DT, and SVR. A lower variance of DT and SVR comes at the cost of a higher bias. For KNN, both bias and variance are larger than in OLS, which is explained by the poor predictive performance of KNN (see Figure \ref{fig:MAPEs_Methods_Boxplot}). For a subsequent analysis of individual treatment effects (ITEs), we choose the least biased estimator that uses latent variables, in our case OLS, despite its higher overall prediction error compared to SVR and DT. 

\subsection{Experiments on Actual Data}
In the following, we analyze ITEs for users with and without smart home devices. The analysis of reduction is carried out with OLS that utilizes an estimate of the HMM latent state because it is found that this method has the lowest bias on semisynthetic data (see Figure \ref{fig:Synthetic_Prediction_Errors}).

\begin{figure*}[hbtp]
\subfloat[]{\includegraphics[width=1.0\textwidth]{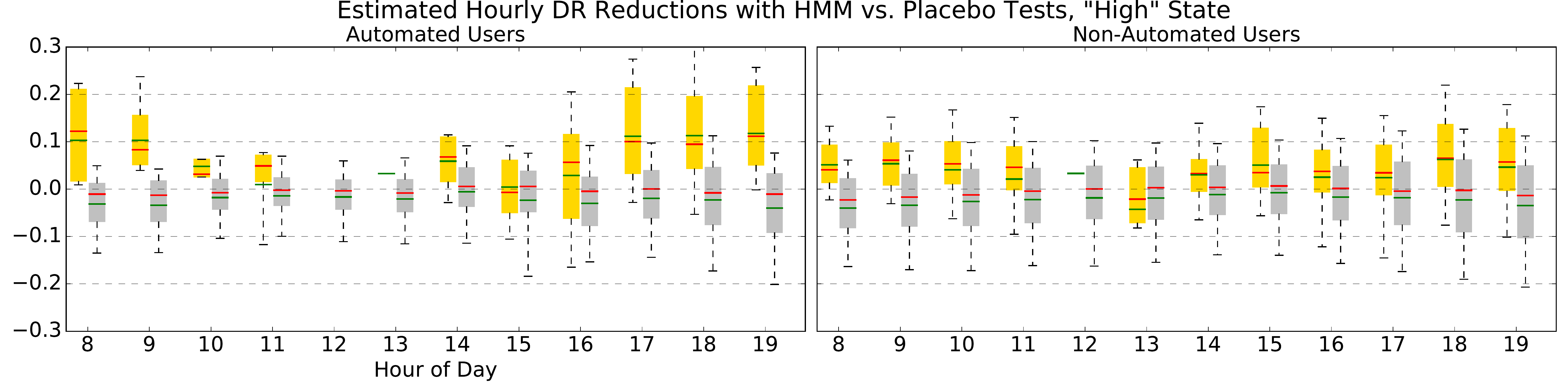}} \hfill \vspace*{-0.45cm}
\subfloat
[]{\includegraphics[width=1.0\textwidth]{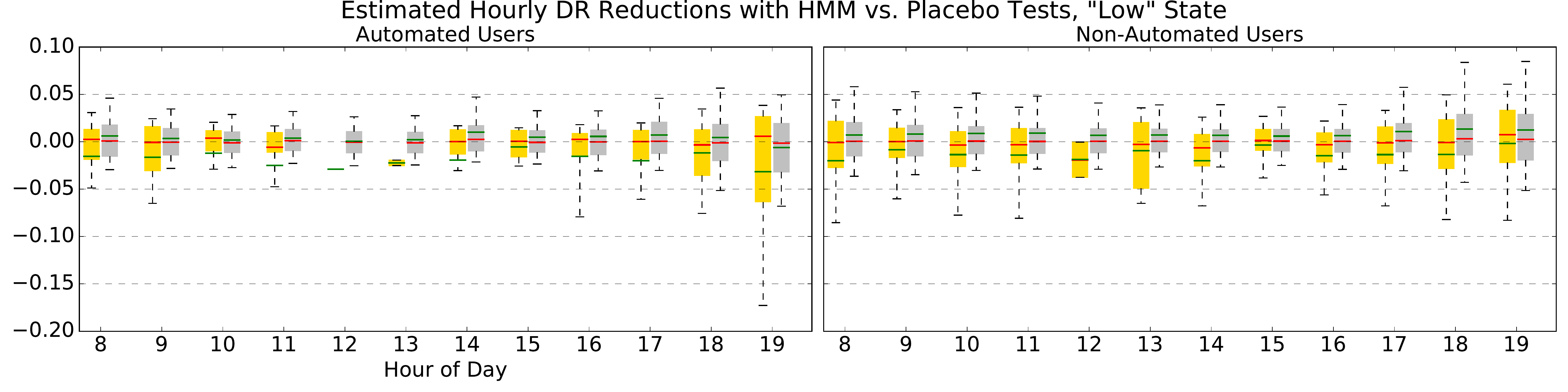}} \hfill \vspace*{-0.35cm}
\caption{Estimated Reduction Across Users by Hour of Day (Yellow) vs. Estimated Reduction for Placebo Events (Gray) for Automated and Non-Automated Users Conditional on Estimated Latent Variable. Red: Median, Green: Mean. Blue Boxes Span 25-75th Percentile, Whiskers 10-90th.}
\label{fig:DR_Reductions}
\vspace*{-0.2cm}
\end{figure*}

In the real data case, only the treatment outcomes $\lbrace y_{i,t}^1 : t \in \mathcal{T} \rbrace$ for user $i$ are observed during DR events, and so the counterfactuals $\lbrace\hat{y}_{i,t}^0 : t \in \mathcal{T} \rbrace$ are predicted to calculate a non-experimental estimate of the DR reduction \eqref{eq:estimated_reduction}. Using Algorithm \ref{alg:RT_prediction} on the pre-signup data $\mathcal{D}_P$ \eqref{eq:pretreatment_data} as training data $\mathcal{D}_{\text{tr}}$ for each user and $\mathcal{D}_{\text{te}} = \mathcal{D}_C \cup \mathcal{D}_T$ \eqref{eq:control_data}, \eqref{eq:treatment_data}, the pointwise reductions across all users and each treatment $t \in \mathcal{T}$ are calculated. Figure \ref{fig:DR_Reductions} shows boxplots of estimated DR reductions conditional on (a) the hour of day, (b) users with and without smart home devices, and (c) the predicted latent states. The gray bars represent ``placebo" events (i.e. a subset of hours $t \in \mathcal{C}$ outside DR treatments hours, but after the signup date) estimated by the same model.

Figure \ref{fig:DR_Reductions} gives rise to two observations: First, the estimated reduction conditional on the ``high" latent state is greater in magnitude for users with smart home devices, following the intuition that the ``high" state describes the operation of smart home devices which can be conveniently shut off during DR hours. In contrast, the lower estimated reductions of regular users during ``high" latent states might reflect the additional hassle cost that incurs for users to manually reduce their consumption. Second, the estimated reductions for both users with and without smart home devices and conditional on the ``low" latent state show mean reductions around zero, contrary to the expectation of a small positive reduction. This might indicate the existence of a threshold representing the standby consumption of users, below which it is hard or impossible to reduce consumption further.

This finding could be particularly meaningful to the DR provider, as it presents a recommendation as to when to call DR events and for which users, which could improve the allocative efficiency of DR targeting and be a stepping stone towards calculating optimal bids.

\section{Conclusion}
\label{sec:Conclusion}
We developed non-experimental estimators from Machine Learning for estimating ITEs of Residential Demand Response and showed that incorporating a latent variable, either with a Conditional Gaussian Mixture Model or a Hidden Markov Model, allows for an improvement in prediction accuracy. This Bayesian approach is motivated by the need to obtain interpretable and physically meaningful results capturing the users' electricity consumption behavior. We then tested the forecasting algorithms on semi-synthetic data to find that Ordinary Least Squares in conjunction with a latent variable produces the least biased estimator for DR reduction. Lastly, this estimator was applied on a residential DR data set to determine hourly reductions of electricity consumption for both users with and without automated electric devices. The highest reductions were found to be among users with home automation devices during ``high" estimated latent states, which in turn provides a recommendation for DR providers for targeting purposes, i.e. to focus on automated users for the highest yield in reduction.

This paper provides only a foundation for more profound analyses in the area of Residential Demand Response. In particular, latent variables can be added as an additional covariate to more computationally demanding estimators, for instance Neural Networks or Random Forests, in order to assess the gain in forecasting precision with latent variables. This is an area to be explored by the established area of STLF, which has traditionally been focusing on maximizing the precision of forecasting algorithms. Further, various extensions to modeling the HMM are worth exploring, such as enlarging the state space of the Markov Chain to enforce a dependency on more than just the previous hour, or increasing the number of hidden states for a given hour (i.e ``low", ``medium", and ``high" consumption). Lastly, the estimated latent variable could be related to a measure of occupancy in residential dwellings, and so a validation of the estimated latent states on ground truth data on occupancy would be interesting if privacy concerns could be overcome.

\section*{Acknowledgment}
We thank Songhwai Oh for interesting discussions.

\bibliographystyle{IEEEtran}
\bibliography{bibliography}


\end{document}